\documentclass[conference]{IEEEtran}
\PassOptionsToPackage{svgnames,dvipsnames}{xcolor}
\IEEEoverridecommandlockouts

\usepackage{cite}
\usepackage{amsmath,amssymb,amsfonts}
\usepackage{algorithm}
\usepackage{algorithmic}

\usepackage{graphicx}
\usepackage{textcomp}
\def\BibTeX{{\rm B\kern-.05em{\sc i\kern-.025em b}\kern-.08em
    T\kern-.1667em\lower.7ex\hbox{E}\kern-.125emX}}

\usepackage{hyperref}
\hypersetup{
    colorlinks=true,
    citecolor=NavyBlue,
    urlcolor=NavyBlue,
    filecolor=NavyBlue,
    linkcolor=NavyBlue
}
\usepackage{multicol}
\usepackage{rotating}
\usepackage{multirow}
\usepackage{booktabs}
\usepackage{array}
\usepackage{makecell}

\usepackage{xspace,xpunctuate}
\usepackage{tabularx}
\usepackage{enumitem}
\usepackage{threeparttable}
\usepackage{svg}
\usepackage{etoolbox}
\usepackage{marvosym}
\AtBeginEnvironment{algorithmic}{\linespread{0.89}\selectfont\small}

\newcommand{\ie}{\textit{i.e.},\xspace}

\newcommand{\eg}{\textit{e.g.},\xspace}

\newcommand{\name}{DataLab}

\newcommand{\edit}[1]{\textcolor{black}{#1}}

\begin{document}

\title{
\name: A Unified Platform for LLM-Powered Business Intelligence
\thanks{This work is done during Luoxuan Weng's internship at Tencent TEG Big Data Platform. We thank the great support from Tencent Cloud ChatBI and Tencent Ad DataWarehouse Team. This work is supported by the National Natural Science Foundation of China (62132017, 62302435, 62421003) and Zhejiang Provincial Natural Science Foundation of China (LD24F020011). Wei Chen and Peng Chen are the corresponding authors.
}
}

\author{
Luoxuan Weng\textsuperscript{† ‡}, Yinghao Tang\textsuperscript{†}, Yingchaojie Feng\textsuperscript{†}, Zhuo Chang\textsuperscript{§ ‡}, Ruiqin Chen\textsuperscript{‡}, Haozhe Feng\textsuperscript{‡},\\
Chen Hou\textsuperscript{‡}, Danqing Huang\textsuperscript{‡}, Yang Li\textsuperscript{‡}, Huaming Rao\textsuperscript{‡}, Haonan Wang\textsuperscript{‡}, Canshi Wei\textsuperscript{‡}, Xiaofeng Yang\textsuperscript{‡},\\
Yuhui Zhang\textsuperscript{‡}, Yifeng Zheng\textsuperscript{‡},
Xiuqi Huang\textsuperscript{†}, Minfeng Zhu\textsuperscript{$\parallel$}, Yuxin Ma\textsuperscript{¶}, Bin Cui\textsuperscript{§}, Peng Chen\textsuperscript{‡ \Letter}, Wei Chen\textsuperscript{† \Letter}
\\
\normalsize
\textsuperscript{†}\textit{State Key Lab of CAD\&CG, Zhejiang University},
\textsuperscript{‡}\textit{Tencent Inc.},
\textsuperscript{$\parallel$}\textit{Zhejiang University},
\\
\normalsize
\textsuperscript{¶}\textit{Southern University of Science and Technology},
\textsuperscript{§}\textit{School of Computer Science, Peking University}
\\
\normalsize
\textsuperscript{†}\{lukeweng, yinghaotang, fycj, huangxiuqi, chenvis\}@zju.edu.cn,
\textsuperscript{‡}\{ruiqinchen, aidenhzfeng, rickhou, daisyqhuang\\
\normalsize
thomasyngli, huamingrao, nanthanwang, caydenwei, felixxfyang, yukiyhzhang, yifengzheng, pengchen\}@tencent.com,\\
\normalsize
\textsuperscript{$\parallel$}minfeng\_zhu@zju.edu.cn,
\textsuperscript{¶}mayx@sustech.edu.cn,
\textsuperscript{§}\{z.chang, bin.cui\}@pku.edu.cn
}

\maketitle

\begin{abstract}
Business intelligence (BI) transforms large volumes of data within modern organizations into actionable insights for informed decision-making. Recently, large language model (LLM)-based agents have streamlined the BI workflow by automatically performing task planning, reasoning, and actions in executable environments based on natural language (NL) queries. However, existing approaches primarily focus on individual BI tasks such as NL2SQL and NL2VIS. The fragmentation of tasks across different data roles and tools lead to inefficiencies and potential errors due to the iterative and collaborative nature of BI. In this paper, we introduce \name, a \textit{unified} BI platform that integrates a \textit{one-stop} LLM-based agent framework with an \textit{augmented} computational notebook interface. \edit{\name \space supports various BI tasks for different data roles in data preparation, analysis, and visualization} by seamlessly combining LLM assistance with user customization within a \textit{single} environment. To achieve this unification, we design a domain knowledge incorporation module tailored for enterprise-specific BI tasks, an inter-agent communication mechanism to facilitate information sharing across the BI workflow, and a cell-based context management strategy to enhance context utilization efficiency in BI notebooks. Extensive experiments demonstrate that \name \space achieves state-of-the-art performance on various BI tasks across popular research benchmarks. Moreover, \name \space maintains high effectiveness and efficiency on real-world datasets from Tencent, achieving up to a \textbf{58.58\%} increase in accuracy and a \textbf{61.65\%} reduction in token cost on enterprise-specific BI tasks.
\end{abstract}


\section{Introduction}
\label{sec:intro}

Business intelligence (BI) aims to transform large volumes of data into actionable insights for informed decision-making~\cite{10.14778/3415478.3415557}. A typical BI workflow includes multiple stages such as data preparation, analysis, and visualization. It requires the collaboration of data engineers, scientists, and analysts using various specialized tools (\eg Visual Studio Code, Power BI, Tableau), which can be highly tedious and time-consuming~\cite{2024-spider2v}. Therefore, modern organizations require advanced techniques to automate and optimize this workflow.


Recent advancements in autonomous agents powered by large language models (LLMs)~\cite{xie2024openagents} offer the potential to streamline the BI workflow. By receiving instructions in natural language (NL), LLM-based agents can perform task planning, reasoning, and actions in executable environments. This can significantly reduce the complexity of many BI tasks, such as code generation~\cite{lai2023ds}, text-to-visualization translation~\cite{10.1145/3654992}, and automated insight discovery~\cite{DBLP:journals/corr/abs-2404-01644}.


However, previous works on LLM-based agents for BI primarily focus on individual tasks or stages without considering the BI workflow as a whole. 
The separation of BI tasks across different data roles and tools impedes seamless information flow and insight exchange, adding to communication costs, delays, and errors~\cite{10035715, meduri2021birecguideddataanalysis}.
For example, data analysts using GUI-based platforms (\eg Power BI) often rely on data engineers working with development tools (\eg PyCharm) to prepare data for analysis or visualization. This reliance necessitates back-and-forth communication between analysts and engineers due to the iterative and collaborative nature of BI~\cite{10.14778/3415478.3415557}. Such procedures highlight the limitations of existing fragmented and fixed agent pipelines~\cite{DBLP:journals/corr/abs-2402-18679}. Consequently, this leads to a significant gap among different roles, tasks, and tools, which hinders timely and informed decision-making.


To bridge this gap, we aim to unify the BI workflow with a \emph{one-stop} LLM-based agent framework in a \emph{single} environment that satisfies the requirements of various data roles.
However, achieving this unification in practical enterprise settings is non-trivial due to the following challenges:


\textbf{C1: Lack of domain knowledge incorporation.}
Existing studies usually leverage clean and synthesized research benchmarks to build and evaluate agents~\cite{su2024tablegpt2largemultimodalmodel}.
However, BI tasks typically involve large and dirty real-world datasets with many ambiguities~\cite{chen2024beaver}.
For example, column names in business data tables may have unclear semantic meanings~\cite{DBLP:journals/corr/abs-2405-00527}, and user queries often include enterprise-specific jargon~\cite{su2024tablegpt2largemultimodalmodel}.
To mitigate these issues, incorporating extensive domain knowledge is essential to enhance agents' understanding of input data and improve their performance on practical BI tasks.
While some approaches adopt fine-tuning~\cite{vm2024fine} or continued pre-training~\cite{ibrahim2024simple} to augment agents' domain-specific capabilities,
acquiring the necessary large and up-to-date training data corpora remains challenging in BI scenarios due to their complexity and dynamic nature.
\edit{Other approaches (\eg Chat2Data~\cite{DBLP:journals/pvldb/ZhaoZL24}) require users to manually integrate domain knowledge through external documents or customized knowledge bases, which is highly inefficient and inconvenient.}

\textbf{C2: Insufficient information sharing across tasks.}
Different tasks are typically managed by corresponding LLM-based agents to achieve optimal performance~\cite{DBLP:journals/corr/abs-2308-08155}. As a complex BI query may encompass multiple tasks, information sharing among the involved agents is critical. For example, the data retrieved by a \textit{SQL writing agent} must be accurately conveyed to a \textit{chart generation agent}.
Therefore, an effective and efficient inter-agent communication mechanism is essential to align their understanding of the overall analysis goals, current data context, and executed actions.
\edit{
However, many existing multi-agent frameworks, such as ChatDev~\cite{qian-etal-2024-chatdev} and CAMEL~\cite{NEURIPS2023_a3621ee9}, rely on unstructured natural language for communication, which can lead to distortions due to the inherent vagueness and redundancy of NL~\cite{hong2023metagpt4}. Consequently, they are inadequate for handling the complexity of BI tasks, which commonly involve diverse information types (\eg data, charts, texts).
}

\textbf{C3: Demand for adaptive LLM context management.}
LLM-based agents depend on their \emph{context windows} (\ie limited input tokens for NL understanding, reasoning, and generation) to complete tasks. Necessary contexts must be provided to ensure a successful and seamless workflow. Meanwhile, in a unified BI platform, vast amounts of multi-modal contexts (\eg code snippets and their execution results, charts and their specifications) are intertwined and often relate to diverse data tables. Obviously, only relevant portions of these contexts are pertinent to specific tasks and should be selectively provided to the agents~\cite{DBLP:conf/icml/ShiCMSDCSZ23}.
\edit{
In contrast, existing works either focus on single-modal contexts tied to specific tasks~\cite{DBLP:conf/icde/RenFHHDHJZYW24} or indiscriminately provide all contexts~\cite{DBLP:journals/corr/abs-2308-08155}, neither of which meets the demands of a unified BI platform.
Thus, adaptive context management tailored for BI scenarios that considers prior states and current user needs is crucial for maintaining system efficiency and cost-effectiveness.
}

In this paper, we introduce \name, a \emph{unified} environment that supports various data tasks throughout the BI workflow, thereby serving different data roles whether they use Markdown, SQL, Python, or no-code, all within a \emph{single} computational notebook. We use notebooks as the foundational system due to their popularity in data science~\cite{10035715} and their iterative nature for the BI workflow~\cite{10.14778/3415478.3415557}.
\name \space adopts an LLM-based agent framework to integrate LLM assistance seamlessly, and a notebook interface to enable user customization flexibly.


To improve agents' performance on enterprise-specific BI tasks (for \textbf{C1}), we develop a \textit{Domain Knowledge Incorporation} module, a systematic approach for automated knowledge generation, organization, and utilization. It leverages data processing scripts (\eg Python code, SQL queries) within the enterprise to extract knowledge associated with databases/tables/columns, thereby uncovering their common usage patterns.
\edit{This module overcomes the practical challenge of manually integrating external knowledge bases for BI.}

To facilitate information sharing across different tasks (for \textbf{C2}), we design an \textit{Inter-Agent Communication} module, a structured mechanism that goes beyond pure NL to enhance the information representation capabilities of agents. It also formulates the information sharing process among agents with a finite state machine (FSM) for a more controlled and efficient flow of communication.
\edit{
This module addresses the challenge of effectively and efficiently sharing multi-modal information generated by various agents throughout the BI workflow.
}

To manage LLM contexts within multi-modal notebooks (for \textbf{C3}), we propose a \textit{Cell-based Context Management} module that represents inter-cell dependencies using directed acyclic graphs (DAGs). These dependency graphs are dynamically updated in response to user modifications, enabling the adaptive selection of pertinent contexts based on specific task requirements.
\edit{
This module resolves the challenge of enhancing agents' context utilization efficiency in BI scenarios.
}

\edit{
Compared to existing works, \name \space stands out due to four benefits: (1) It delivers satisfactory performance across various BI tasks, rather than focusing on individual tasks; (2) It is well-suited for real-world applications, not just research benchmarks; (3) It offers a unified platform to satisfy different user requirements, rather than catering
to a single data role; and (4) It integrates LLM assistance with user customization, instead of relying solely on
end-to-end result generation.
}

In summary, the major contributions of our work are:

\begin{itemize}
    \item We present \name, a platform that unifies the BI workflow with the integration of a one-stop LLM-based agent framework and a computational notebook interface, to bridge the gap among different roles, tasks, and tools.
    \item We develop a systematic approach for domain knowledge incorporation to enhance LLM-based agents' performance on enterprise-specific BI tasks in practical settings.
    \item We introduce a structured communication mechanism to formulate the information sharing process among different agents to facilitate their cross-task performance.
    \item We propose an adaptive context management strategy to improve agents' context utilization abilities within computational notebooks for efficiency and cost-effectiveness.
    \item We extensively evaluate \name \space on both research benchmarks and real-world business datasets from Tencent, demonstrating its performance on various BI tasks. \edit{We also showcase the practical applications of \name \space at Tencent TEG's Big Data Platform through user feedback.}
\end{itemize}
\begin{figure*}[ht]
  \centering
  \includegraphics[width=\textwidth]{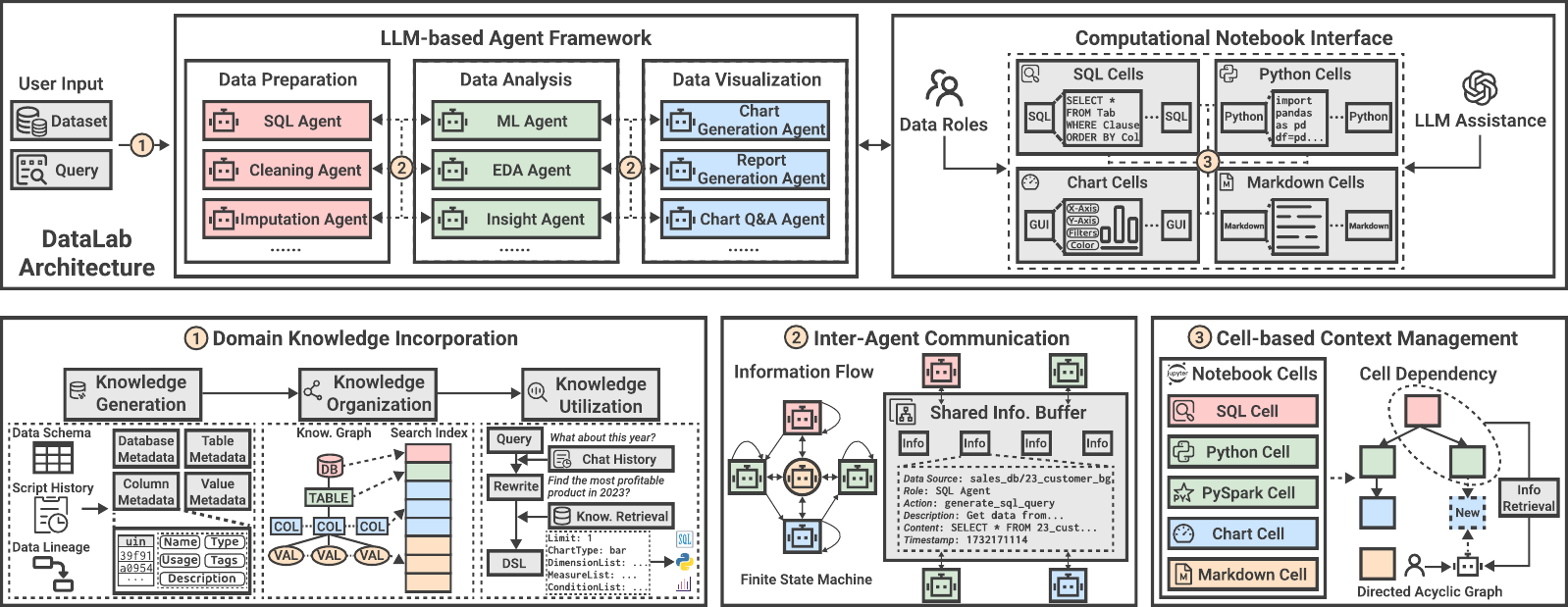}
  \vspace{-7mm}
  \caption{Overview of \name \space and its three critical modules.}
  \label{fig:overview}
  \vspace{-7mm}
\end{figure*}

\section{Background}


\subsection{BI Workflow}

\edit{
The BI workflow traditionally includes several key stages: data collection, storage, preparation, analysis, and visualization. Data Collection and Storage establish the foundation, typically supported by cloud infrastructures (\eg Tencent TCHouse-C, AWS) and specialized data applications (\eg BigQuery, Airflow) with mature ecosystem integration. Attempting to replicate these practices in alternative environments could introduce inefficiencies and anti-patterns. Moreover, the core value of modern BI lies in transforming \textit{raw data} into actionable \textit{insights} through data preparation, analysis, and visualization~\cite{10.14778/3415478.3415557, meduri2021birecguideddataanalysis}. Therefore, our work primarily focuses on the latter three stages, emphasizing \textit{interoperability} with existing tools rather than reinventing them. For data collection and storage, we provide plugin APIs to enable seamless integration with third-party data connectors, ensuring flexibility and compatibility within the broader BI ecosystem.
}

\textbf{Data Preparation}~\cite{DBLP:conf/sigmod/ChenLJSFF0024} ensures data consistency, correctness, and quality. This usually includes cleaning, structuring, and enriching raw data into a format suitable for further analysis. Following preparation, \textbf{Data Analysis}~\cite{DBLP:conf/icde/ZhuCNNXHWMWZTL24} applies statistical and analytical techniques to extract insights, aiming to uncover patterns, trends, and correlations. Finally, \textbf{Data Visualization}~\cite{10.1145/3654992} presents analyzed data in visual formats like charts, graphs, and dashboards, which makes complex data easier to understand and interpret for decision-makers.

The data roles involved in the BI workflow are specific to different organizations. Among them, data engineers, scientists, and analysts are usually indispensable. \textbf{Data Engineers} are primarily tasked with data preparation, constructing and administering data pipelines to ensure that data is accurately cleansed and structured for subsequent analysis. They typically use SQL and Python for data processing, and use cloud computing platforms like AWS for data storage and ingestion. \textbf{Data Scientists} engage in data preparation and analysis, applying advanced statistical and machine learning methodologies to extract insights and forecast trends from intricate datasets. They are familiar with Python/R and popular data science libraries like \texttt{Pandas}. \textbf{Data Analysts} concentrate on data analysis and visualization, analyzing data to discern patterns and conveying findings through detailed visual reports and dashboards. They use SQL to query data, and rely on BI platforms like Tableau to perform and share their analyses.

In modern enterprises, a complex BI workflow requires the collaboration of multiple data roles across various stages. The current fragmentation of tools for data preparation, analysis, and visualization introduces frictions and delays in timely decision-making. Therefore, an integrated and unified platform can serve as a shared environment for distinct user groups, facilitating the efficiency, transparency, and productivity of BI.

\subsection{LLM-based Agents for BI}
\label{subsec: bg-multiagents}
LLM-based agents are autonomous systems powered by LLMs that can perceive environments, execute tasks, make decisions, and interact with users in complex contexts~\cite{xie2024openagents}. These agents comprise profiling, memory, planning, and action modules, which respectively define the agent's role, facilitate operations in dynamic environments through recall and future action planning, and convert decisions into outputs~\cite{wang2024survey}. In BI scenarios, agents receive users' NL queries and then perform data preparation, analysis, and visualization.
\edit{
By interpreting execution results, they can complete many BI tasks.
For example, data preparation involves tasks like \textbf{NL2SQL}~\cite{10.14778/3641204.3641221} and \textbf{NL2DSCode}~\cite{lai2023ds}, while data analysis and visualization rely on \textbf{NL2Insight}~\cite{sahu2024insightbench} and \textbf{NL2VIS}~\cite{10.1145/3654992}, respectively.
}

However, most existing LLM-based agents are limited to individual tasks and do not meet the diverse user requirements of a complex BI workflow. Moreover, they often neglect the integration of enterprise-specific knowledge, resulting in unsatisfactory performance on proprietary business datasets. This lack of generalizability and customizability highlights the need for a structured and adaptive agent framework for BI.
\section{Overview}
\label{sec:overview}

\begin{figure}[t]
    \centering
    \includegraphics[width=\linewidth]{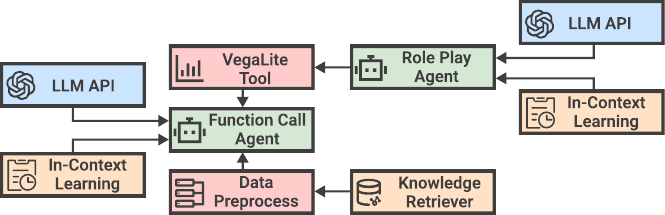}
    \vspace{-7mm}
    \caption{Example agent workflow for NL2VIS.}
    \label{fig:nl2vis}
    \vspace{-7mm}
\end{figure}
\begin{figure}[t]
    \centering
    \includegraphics[width=\linewidth]{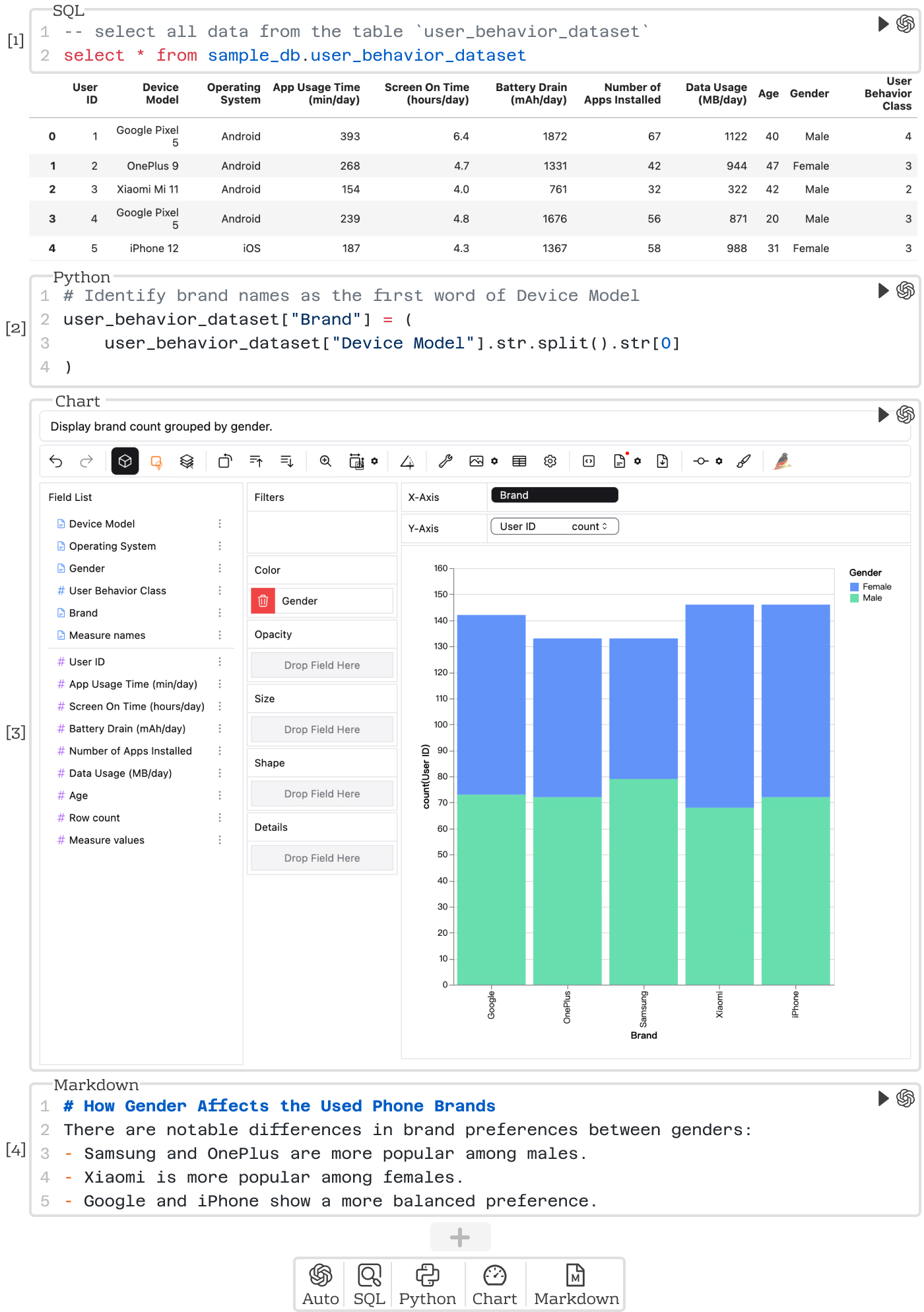}
    \vspace{-7mm}
    \caption{The notebook interface of \name.}
    \label{fig:notebook}
    \vspace{-7mm}
\end{figure}

\textbf{Architecture Overview.} As illustrated in Figure~\ref{fig:overview}, \name \space consists of two primary components: (1) \textit{LLM-based Agent Framework} and (2) \textit{Computational Notebook Interface}.
\begin{itemize}[leftmargin=10pt]
    \item \textbf{LLM-based Agent Framework.}
    In \name, multiple agents are designed for different BI tasks based on user requirements.
    To achieve this, we first identify several common BI procedures and abstract them into \emph{data tools} that can be called upon by agents during inference. Example tools include a Python sandbox for code execution and a Vega-Lite environment for visualization rendering. Accompanied by other auxiliary components like memory modules, each BI agent is represented as a DAG for high flexibility and easy extensibility. Within the DAG, nodes depict reusable components (\eg LLM APIs, tools) and edges depict their connections (\eg file transfer across tools). Figure~\ref{fig:nl2vis} illustrates an example agent workflow for NL2VIS.  Additionally, we add a \emph{proxy agent} to the framework, which serves as a hub to directly interact with users and allocate tasks to each specialized agent based on user queries. \edit{Compared to existing approaches that primarily focus on individual tasks (\eg NL2VIS~\cite{10121440}, NL2SQL~\cite{DBLP:conf/icde/RenFHHDHJZYW24}), \name's agent framework supports a wide array of tasks for various data roles across the BI workflow through multi-agent collaboration.}
    \item \textbf{Computational Notebook Interface.}
    \name's notebook interface (Figure~\ref{fig:notebook}) serves as a unified, interactive, and collaborative environment for different data roles to complete their specialized tasks. To achieve this, we augment JupyterLab (a widely used notebook interface) to support (1) multi-language cells and (2) on-the-fly LLM assistance. First, \name \space wrangles SQL, Python/PySpark, Markdown, and Chart cells altogether, allowing both technical and non-technical users to easily adopt their familiar workflows on the notebook. Going beyond traditional notebooks that only support Python and Markdown, \name \space notebooks directly connect to backend databases for SQL query execution, and feature GUI-based dashboards~\cite{DBLP:journals/corr/abs-2406-11637} similar to Tableau for visualization authoring. Second, we integrate our LLM-based agent framework seamlessly into each notebook cell. Users can get LLM assistance both at notebook- and cell-level. Specifically, users toggle an input box and type their analytic queries, which are then processed by the agents in our framework. These agents can create new cells or modify existing ones in the notebook. Users can subsequently examine the results and make further customizations flexibly.
    \edit{
    Compared to existing end-to-end approaches (\eg CHESS~\cite{DBLP:journals/corr/abs-2405-16755}, LIDA~\cite{dibia-2023-lida}), \name's notebook interface enables flexible user intervention to adapt LLM-generated results to real-world BI scenarios.
    }
\end{itemize}

\textbf{DataLab Workflow.} Upon receiving an NL query and the associated dataset, \name \space analyzes the dataset and interprets the query, incorporating domain knowledge (\edit{Figure~\ref{fig:overview}}\includegraphics[width=1em]{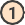}) before feeding them into LLMs. Then, \name \space leverages various agents to complete the task, which may involve information sharing with each other through a structured communication mechanism (\edit{Figure~\ref{fig:overview}}\includegraphics[width=1em]{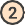}). Subsequently, the corresponding result will be presented in the notebook. Users can either accept, edit, or reject the result and continues the BI workflow. Meanwhile, a context management strategy (\edit{Figure~\ref{fig:overview}}\includegraphics[width=1em]{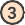}) automatically generates and maintains cell dependencies within the notebook to promote further agent calls. Next, we provide an overview of \name's three critical modules.

\begin{itemize}[leftmargin=10pt]
    \vspace{-1mm}
    \item \textbf{Domain Knowledge Incorporation.} This module takes a data table's schema, its associated script history (\eg SQL queries, Python code), and its data lineage information as input. Specifically, the schema provides a basic overview of the table and its columns, including their names and types. The associated data processing scripts, which are created by professionals and executed every day within the organization, reflect the semantic meanings and common usage patterns of the table and its columns. And the data lineage information~\cite{8731450}, which reveals interrelationships among distinct tables and columns across the organization, can serve as an auxiliary resource for domain knowledge extraction.
    Based on the input, the module leverages LLMs to automatically generate the \emph{knowledge components} (\eg descriptions, usages) of databases, tables, columns, and certain values. These knowledge components are then organized in a knowledge graph to facilitate further retrieval and utilization, which translate ambiguous user queries into structured domain-specific languages (DSLs) for improved agent performance on enterprise-specific BI tasks.
    \item \textbf{Inter-Agent Communication.} This module formulates the information flow process among different agents as an FSM to enable more control over their communications, with nodes representing agents and edges representing inter-agent information transition directions. Upon task completion, each agent's outputs are formatted into structured \emph{information units}~\cite{hong2023metagpt4}, comprising key characteristics such as the associated table's identifier and a concise description of the actions performed. The module also maintains a shared information buffer to proactively exchange information based on the FSM to improve communication efficiency.
    \item \textbf{Cell-based Context Management.} This module identifies cell dependencies within a notebook based on variable references and constructs a DAG, where nodes represent cells and edges denote their dependencies. Notably, data variables in Python or SQL cells are meticulously tracked, such as \texttt{DataFrames} and \texttt{SELECTs}. Given a user query, the module traverses the DAG to locate relevant cells, performs pruning based on task types, and retrieves information from the shared buffer. Then, the original cells and their corresponding information units are fed to the proxy agent as necessary contexts to facilitate task completion.
\end{itemize}

\section{Domain Knowledge Incorporation}
\label{sec:dk}

In this section, we introduce \name's \textit{Domain Knowledge Incorporation} module, which encompasses three stages, namely knowledge generation, organization, and utilization.

\subsection{Knowledge Generation}
\label{sec:know_gen}

Ambiguities are pervasive in real-world BI scenarios, manifesting both in the underlying databases and users' NL queries. For example, consider the query, \textit{`show me the income of TencentBI this year'}, which involves three columns: \textit{`prod\_class4\_name'}, \textit{`shouldincome\_after'}, and \textit{`ftime'}. The semantic relationships between these column names and the user's request are often vague, leading to LLMs' suboptimal performance on such tasks. To mitigate these issues, existing approaches integrate table schema~\cite{10.1145/3654992} into prompts and adopt retrieval-augmented generation (RAG)~\cite{su2024tablegpt2largemultimodalmodel} to improve LLMs' domain-specific capabilities. We categorize three types of domain knowledge commonly utilized for BI tasks:
\begin{itemize}
    \vspace{-1mm}
    \item \textbf{Metadata:} Information about data structure and attributes, such as table and column names, types, descriptions, and common usage patterns.
    \item \textbf{Business Logic:} Rules and processes that dictate how data is used and interpreted within the business.
    \item \textbf{Jargon:} Specialized terminologies and acronyms specific to the industry or organization.
\end{itemize}

\vspace{-1mm}
\edit{
Many existing works assume that such knowledge can be manually constructed and integrated by domain experts. However, in large organizations, enormous numbers of data tables are commonly involved, making manual curation and maintenance highly impractical.
To overcome this limitation, we first conducted an extensive investigation at Tencent.
It was observed that,} while 85\% of the tables lack comprehensive metadata, they are predominantly linked to various SQL or Python scripts utilized for data processing. These scripts reveal common usage patterns within practical business contexts. Additionally, for those tables lacking adequate processing scripts, data lineage information, which elucidates their connections to other tables or columns throughout the organization, provides an alternative resource for metadata imputation. Therefore, inspired by LLMs' exceptional code understanding and reasoning abilities, we propose an LLM-based knowledge generation approach (Algorithm~\ref{alg:metadata}) that leverages script history to abstract and summarize knowledge components through meticulously-designed prompting techniques. This automated approach comprises a Map-Reduce process with a self-calibration mechanism~\cite{DBLP:conf/emnlp/TianMZSRYFM23} to generate high-quality knowledge for databases, tables, and columns.

\vspace{-2mm}
\begin{algorithm}[]
\caption{LLM-based Knowledge Generation}
\label{alg:metadata}
\begin{algorithmic}[1]
\REQUIRE Schema $\mathcal{S}$, Script History $\mathcal{H}$, \\
\quad \, Lineage Information $\mathcal{L}$, Score Threshold $\mathcal{T}$
\ENSURE Database/Table/Column Knowledge $\mathcal{D}$, $\mathcal{T}$, $\mathcal{C}$
\STATE $\mathcal{H} \gets \text{preprocess}(\mathcal{H})$ \textcolor{NavyBlue}{\textit{// Duplicated/Similar script filtering}}
\STATE $\text{map\_res} \gets [\ ]$
\FOR{\textbf{each} historical script $h_i \in \mathcal{H}$}
    \WHILE{$s_i < \mathcal{T}$}
        \STATE $d_i, t_i, c_i \gets \text{LLM}(h_i, \mathcal{S}, \mathcal{L})$ \textcolor{NavyBlue}{\textit{// Knowledge generation}}
        \STATE $s_i \gets \text{LLM}(d_i, t_i, c_i)$ \textcolor{NavyBlue}{\textit{// Self-calibration}}
    \ENDWHILE
    \STATE $\text{map\_res}.\text{append}([d_i, t_i, c_i])$
\ENDFOR
\STATE $\mathcal{D}, \mathcal{T}, \mathcal{C} \gets \text{LLM}(\text{map\_res}, \mathcal{S}, \mathcal{L})$ \textcolor{NavyBlue}{\textit{// Knowledge synthesis}}
\RETURN $\mathcal{D}, \mathcal{T}, \mathcal{C}$
\end{algorithmic}
\end{algorithm}
\vspace{-3mm}

\textbf{Knowledge Components.}
Considering the previously defined knowledge categories, metadata and business logic can be deduced from data processing scripts, as both SQL queries and Python code support data manipulation operations like filtering and aggregation. Business logic is essential for computing \textit{derived columns} which, though absent in the original table, hold significant value in business contexts. In contrast, jargon primarily exists in user queries or organization wikis (\ie documents), necessitating enterprise-specific glossaries for management and application. The \textit{knowledge components} that our automated approach can generate are outlined below:
\begin{itemize}
    \item \textbf{Database Level:} \textit{description, usage, tags.}
    \item \textbf{Table Level:} \textit{description, usage, organization, key column names, key derived attribute names, tags.}
    \item \textbf{Column Level:} \textit{description, usage, type, tags, derived column information (name, description, usage, calculation logic, related columns, tags).}
\end{itemize}

These knowledge components are structured using JSON formats to improve LLMs' generation performance.

\textbf{Map Phase.}
Given a data table, we take its schema $\mathcal{S}$, its script history $\mathcal{H}$, and its lineage information $\mathcal{L}$ as input. During the map phase, each distinct historical script $h_i$ is individually processed using an LLM as the mapping model to produce corresponding knowledge components. The LLM is prompted to carefully analyze the script's semantic content and logical structure, aiming to extract critical information relevant to the specific business context. To mitigate LLMs' hallucination issues, focus is restricted to the involved databases, tables, and columns. This process results in the generation of a set of knowledge components associated with the script $h_i$.

\textbf{Self-Calibration.}
Within each iteration of the map phase, we integrate a self-calibration mechanism that leverages LLMs' self-reflection abilities~\cite{DBLP:conf/emnlp/JiYXLIF23} to evaluate the intermediate results using a numerical score ranging from 1 to 5. Specifically, we instruct the LLM to consider multiple aspects of the knowledge components (\eg correctness, comprehensiveness, clarity) and provide several manually crafted in-context examples to demonstrate the scoring criteria. Should the rating score $s_i$ fall below the predefined threshold $\mathcal{T}$, the knowledge generation process must be repeated. Therefore, this feedback loop ensures the generation quality of each iteration.

\textbf{Reduce Phase.}
During the reduce phase, we aim to synthesize the individual results derived from each historical script to produce the final sets of knowledge components $\mathcal{D}$, $\mathcal{T}$, and $\mathcal{C}$ for the involved database, table, and columns, respectively. The LLM is instructed to meticulously scrutinize, aggregate, and summarize the information from all separate results to ensure a consistent and conflict-free collective result.

For each data table at Tencent, we execute the above Map-Reduce process to generate a comprehensive and high-quality set of knowledge components, which can significantly benefit many downstream BI tasks.

\vspace{-1mm}
\subsection{Knowledge Organization}

We employ a knowledge graph $\mathcal{G}=(\mathcal{V}, \mathcal{E})$ to systematically organize the knowledge generated by our automated approach (\ie metadata and business logic) and the manually crafted enterprise-specific glossaries (\ie jargon).

\begin{figure}[t]
    \centering
    \includegraphics[width=\linewidth]{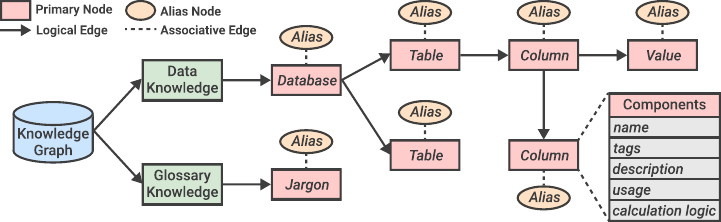}
    \vspace{-7mm}
    \caption{Structure of the knowledge graph.}
    \label{fig:kg}
    \vspace{-5mm}
\end{figure}

As depicted in Figure~\ref{fig:kg}, the knowledge graph adopts a tree-based structure for knowledge organization. The nodes $\{\mathcal{V}\}$ are structured into five primary types: \textit{database}, \textit{table}, \textit{column}, \textit{value}, and \textit{jargon}, each comprising various components (\eg \textit{description}, \textit{usage}) and uniquely identified by a \textit{name}. To address the common challenge of terminological inconsistencies in user queries (\eg synonyms, acronyms), an additional node type, \textit{alias}, has been introduced. This node type contains alternative terms associated with the official \textit{name} of other node types. \edit{It is predetermined based on enterprise-specific glossaries and can be dynamically updated in real-world applications}. The relationships between these nodes are represented by edges $\{\mathcal{E}\}$, which delineate both \textit{logical relationships} among the primary node types and \textit{associative relationships} between \textit{alias} nodes and other primary nodes.

To facilitate efficient knowledge retrieval, we develop a task-aware indexing mechanism for graph nodes, utilizing Elasticsearch~\cite{gormley2015elasticsearch} for full-text search and StarRocks~\cite{starrocks} for embedding search. This supports both lexical and semantic matching of knowledge nodes in response to user queries. The indexing structure is designed as triplets ($\{\text{name}, \text{content}, \text{tag}\}$), where the $content$ field is a concatenation of knowledge components specified based on the various requirements of downstream tasks. For instance, some tasks necessitate the \textit{calculation logics} while others only need \textit{descriptions} for successful completion. By dynamically selecting the appropriate index, we ensure that knowledge retrieval is both efficient and effective.

\subsection{Knowledge Utilization}
\label{sec:know_utilize}

\vspace{-3mm}
\begin{algorithm}[H]
\caption{Knowledge Retrieval}
\label{alg:know_retrieve}
\begin{algorithmic}[1]
\REQUIRE User Query $\mathcal{Q}$, Knowledge Graph $\mathcal{G}$
\ENSURE Knowledge Nodes $\mathcal{V_Q}$
\STATE $\mathcal{V_Q} \gets \emptyset$
\STATE \textbf{Coarse-Grained Retrieval:}
\STATE $\mathcal{V_Q} \gets \text{lex\_search}(\mathcal{Q}, \mathcal{G}) + \text{sem\_search}(\mathcal{Q}, \mathcal{G})$
\STATE \textbf{Fine-Grained Ordering:}
\FOR{\textbf{each} node $v_i \in \mathcal{V_Q}$}
    \IF{$v_i.\text{type} == \text{alias}$}        
        \STATE $v_i \gets \text{backtrack}(v_i)$ \textcolor{NavyBlue}{\textit{// Backtrack to a primary node}}
    \ENDIF
    \STATE \textcolor{NavyBlue}{\textit{// Compute a weighted matching score}}
    \STATE $\text{score}_i \gets \omega_1 \cdot \text{lex\_eval}(\mathcal{Q}, v_i) + \omega_2 \cdot \text{sem\_eval}(\mathcal{Q}, v_i) +$ \\
           \qquad \quad \,\,\,\,\, $\omega_3 \cdot \text{LLM\_eval}(\mathcal{Q}, v_i)$
\ENDFOR
\STATE $\mathcal{V_Q}.\text{sort}(\text{score}_i)$ \textcolor{NavyBlue}{\textit{// Rank by matching score}}
\RETURN $\mathcal{V_Q}.\text{topK}$
\end{algorithmic}
\end{algorithm}
\vspace{-3mm}

As shown in Figure~\ref{fig:overview}\includegraphics[width=1em]{figs/icons/id_1.pdf}, given a user query $\mathcal{Q}$, we initially rewrite it to enhance clarity and detail. We then retrieve its relevant knowledge from the knowledge graph $\mathcal{G}$. Following this, the query is translated into a DSL specification, facilitating downstream tasks like NL2SQL and NL2VIS.

\textbf{Query Rewrite.}
In addition to ambiguities, user queries can also be incomplete or underspecified, especially in multi-round interactions. For example, queries might omit prior context with phrases like \textit{`what about'}. To ensure effective knowledge retrieval, the original query is enhanced and rewritten into a clearer and more detailed form, incorporating relevant prior information when available. Notably, temporal references (\eg \textit{`last year'}) are also standardized based on the current time.

\textbf{Knowledge Retrieval.}
To enhance LLMs' domain specific performance by integrating relevant knowledge into their context alongside the query, the selection and ordering of knowledge nodes from the knowledge graph are crucial. We employ a coarse-to-fine approach (Algorithm~\ref{alg:know_retrieve}) to ensure comprehensive and precise knowledge retrieval.

\begin{itemize}[leftmargin=10pt]
    \item \textbf{Coarse-Grained Retrieval:} We perform lexical and semantic searches to retrieve a coarse set of knowledge nodes via token matching and embedding similarity between the query and each node's indexing triplet. We set a rather loose threshold to maximize recall. For \textit{alias} nodes, we trace back to identify the nearest primary nodes (\ie database/table/column/value/jargon nodes).
    \item \textbf{Fine-Grained Ordering:} To prioritize the retrieved nodes, we implement a scoring mechanism that calculates a weighted matching score for each node, assessing its relevance to the query. This involves a three-stage evaluation: token-based (\ie lexicon), embedding-based (\ie semantics), and LLM-based (\ie overall relevance)~\cite{DBLP:conf/acl/ChiangL23}. Each stage yields a normalized score, with specific calculation methods and weights tailored to different BI tasks. The final set of knowledge nodes $\mathcal{V_Q}$ is determined by sorting the initial node set according to these scores and selecting the top-$K$ nodes, where $K$ is set to a relatively large value to ensure a comprehensive coverage.
\end{itemize}

\textbf{DSL Translation.}
The final step translates the query into a DSL specification, a common routine in BI scenarios~\cite{su2024tablegpt2largemultimodalmodel}. This JSON structure specifies the relevant data and processing requirements, including fields such as \textit{`MeasureList'} (\ie numerical columns), \textit{`DimensionList'} (\ie categorical columns), and \textit{`ConditionList'} (\ie filters). We prompt an LLM for DSL translation, providing detailed instructions and in-context examples to improve its performance. The generated DSL specification is validated using JSON Schema~\cite{DBLP:conf/www/PezoaRSUV16} to ensure syntactic and semantic correctness. This specification can then be directly converted to high-level languages like SQL and Vega-Lite based on predefined rules, or used to enhance free-form code generation for complex tasks like NL2Insight, thereby facilitating LLMs' performance in business settings.

We also introduce a fallback strategy to address scenarios where relevant knowledge is scarce, especially for in-the-wild tables. Specifically, we develop a \textbf{Data Profiling} module that systematically extracts information from the table. This module consists of two stages: (1) heuristics-based analysis, which identifies and calculates each column's name, data type (\eg \textit{float}, \textit{string}), basic statistics (\eg \textit{min}, \textit{max}), and a random sample list, and (2) LLM-based interpretation, which feeds the extracted information to an LLM to generate a semantic description of each column and the overall table. Together, these stages produce a comprehensive summary of the table, aiding in the DSL translation process.
\section{Inter-Agent Communication}
\label{sec:com}

In this section, we introduce \name's \textit{Inter-Agent Communication} module, which facilitates efficient communication among multiple agents to complete complex BI tasks.
\edit{These agents are created and optimized beforehand through DAG-based workflows (Section~\ref{sec:overview}) to meet different user needs. For example, a \textit{SQL writing agent} is specialized for NL2SQL tasks and a \textit{chart generation agent} is for NL2VIS tasks.}

\begin{figure}[t]
    \centering
    \includegraphics[width=\linewidth]{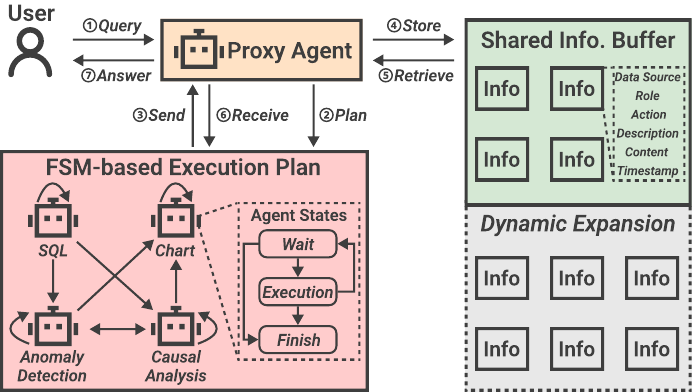}
    \caption{Workflow of \textit{Inter-Agent Communication}.}
    \label{fig:fsm}
    \vspace{-7mm}
\end{figure}

\textbf{Workflow.} As shown in Figure~\ref{fig:fsm}, upon receiving a user query, the proxy agent initiates an analysis to formulate an execution plan (defined by an FSM), which comprises multiple subtasks allocated to various agents (\textit{Steps 1-2}). It then dynamically manages the communication among involved agents based on task progression by retrieving information from a shared buffer and forwarding it to the agents to support subtask execution (\textit{Steps 5-6}). Upon completion of the subtasks, the proxy agent stores the agents' outputs in the buffer (\textit{Steps 3-4}). Finally, once all subtasks are completed, the proxy agent generates a final answer and returns it to the user (\textit{Step 7}).

\textbf{Information Format Structure.} A critical consideration in multi-agent collaboration is \textit{what `language' agents use to communicate}. In BI scenarios, the information exchanged among agents is diverse, encompassing types such as SQL queries, Python code, and charts.
\edit{
This variety poses a significant challenge in ensuring accurate information sharing without introducing redundancy or miscommunication.
Existing frameworks~\cite{qian-etal-2024-chatdev, NEURIPS2023_a3621ee9} that rely on unstructured natural language for communication suffer from vagueness and inefficiencies.
}
To address this, we design a structured information format~\cite{hong2023metagpt4} tailored to unique characteristics of BI scenarios.

An \textit{information unit} comprises six fields: \textit{Data Source}, \textit{Role}, \textit{Action},  \textit{Description}, \textit{Content}, and \textit{Timestamp}. \textit{Data Source} identifies the dataset manipulated by the agent (\eg a table identifier). \textit{Role} indicates the identity of the agent (\eg SQL Agent). \textit{Action} specifies the agent's behavior (\eg \texttt{generate\_sql\_query}). \textit{Description} provides a summary of the agent's executed actions for the task (\eg writing a SQL query to extract specific data from the table). \textit{Content} details the agent's output (\eg a generated SQL query). \textit{Timestamp} records the completion time of the task. To maintain consistent communication, all agents produce messages in this format.

\textbf{Information Sharing Protocol.} \edit{Another key challenge to resolve is \textit{how to ensure efficient information sharing among agents}.} One extreme approach is that each agent receives information solely from its predecessor based on a plan chain~\cite{DBLP:conf/nips/Wei0SBIXCLZ22}. While this minimizes data volume, it may cause the agent overlooking essential context (\eg background information pertinent to the task). Conversely, allowing any agent unrestricted access to the information of all others is inefficient, as irrelevant context can degrade LLMs' reasoning quality~\cite{DBLP:conf/emnlp/self_polish} and introduces additional computational overhead~\cite{li2024mixloraenhancinglargelanguage} during inference. Therefore, each agent should only receive the most relevant information for task completion.

To this end, we introduce two mechanisms to achieve this:

\begin{itemize}[leftmargin=10pt]
    \item \textbf{Shared Information Buffer} is a location for agents to store and retrieve information. Upon task completion, an agent deposits the produced information into this buffer via the proxy agent. This mechanism decouples information producers from consumers, thereby reducing synchronization overhead. Consequently, inter-agent communication becomes asynchronous and non-blocking, allowing producers to continue processing without awaiting the retrieval of information by consumers, and vice versa. On the other hand, to manage the substantial information volumes often associated with multi-agent collaboration, we employ a dynamically growing buffering mechanism. Specifically, when the buffer reaches capacity, it automatically increases its size (\ie doubling the current capacity). Additionally, outdated information is periodically cleared. For example, if an agent's information is updated based on execution feedback, the original information is removed. This ensures that the buffer can maintain high performance while efficiently adapting to changing workloads.
\item \textbf{Selective Retrieval} determines the information an agent receives from others. Inspired by the message-passing mechanism in TCP/IP~\cite{tcp}, we design an FSM-based approach to implement this. Specifically, the proxy agent analyzes the user query and generates an FSM based on task requirements and each agent's abilities to orchestrate multi-agent information sharing, with nodes depicting agents and edges depicting information transition directions. Each agent operates within three states: \textit{Wait, Execution, \textnormal{and} Finish}. When an agent (\ie acting as a \textit{`client'}) needs to execute a subtask, the proxy agent (\ie acting as a \textit{`server'}) selects its relevant information from the shared buffer based on the FSM, and forwards it to the client agent. Upon receipt, the agent transitions from the \textit{Wait} state to the \textit{Execution} state, performs the necessary action, and produces structured information, which is then sent back to the proxy agent. The proxy agent, upon receiving the response, stores it in the buffer, and the client agent reverts to the \textit{Wait} state. This loop continues until all subtasks are completed, at which point all agents transition to the \textit{Finish} state and release their resources.
\end{itemize}
\section{Cell-based Context Management}


In this section, we introduce \name's \textit{Cell-based Context Management} module, which adaptively manages contexts in a notebook to ensure system efficiency and cost-effectiveness.

Thanks to the \textit{Inter-Agent Communication} module (Section~\ref{sec:com}), \name \space can handle complex BI tasks that require the collaboration of multiple agents. Specifically, user queries will trigger the presentation or modification of cells in the notebook, each corresponding to specific information units in the shared buffer. However, the previous module primarily aims at facilitating \textit{individual} task completion for \textit{single} data roles. In contrast, real-world BI scenarios often involve \textit{multiple} data roles working on \textit{different} tasks and collaborating within a unified notebook. This typically results in a multitude of multi-language cells (\eg SQL, Python, Chart) generated or altered by either agents or users themselves. When addressing a new user query, it is crucial to efficiently provide the proxy agent with the necessary contexts from the notebook. Simply supplying all cells and their associated information units is impractical due to inefficiency and high token costs. Therefore, we aim to identify the \textit{minimum set} of relevant cells to minimize token usage without compromising agent performance.
\edit{
However, notebooks are inherently dynamic and collaborative environments, where iterative modifications reshape contextual relevance, and multi-role contributions create fragmented contexts requiring task-specific filtering.
}
To address this challenge, we model cell dependencies within the notebook as a DAG based on variable references, and propose an adaptive context retrieval mechanism.

\vspace{-2mm}
\begin{algorithm}
\caption{DAG Construction}
\label{alg:dag}
\begin{algorithmic}[1]
\REQUIRE Notebook Cells $\mathcal{C}$
\ENSURE Dependency DAG $\mathcal{G}$
\STATE $\text{v\_hash}, \text{cell\_refs} \gets \emptyset, \emptyset$
\STATE \textcolor{NavyBlue}{\textit{// Identify new variables in each cell}}
\FOR{\textbf{each} cell $c \in \mathcal{C}$}
    \IF{$c.\text{type} == \text{Python}$}        
        \STATE $\text{ast} \gets \text{construct\_ast}(c)$
        \STATE $\text{new\_v} \gets \text{find\_global\_variables}(\text{ast})$
        \STATE $\text{v\_hash}[\text{new\_v}] \gets c$
    \ELSIF{$c.\text{type} == \text{SQL}$}
        \STATE $\text{data\_v} \gets \text{find\_data\_variable}(c)$
        \STATE $\text{v\_hash}[\text{data\_v}] \gets c$
    \ENDIF
\ENDFOR
\STATE \textcolor{NavyBlue}{\textit{// Find referenced cells for each cell}}
\FOR{\textbf{each} cell $c \in \mathcal{C}$}
    \STATE $\text{external\_v} \gets \text{find\_external\_variables}(c)$
    \STATE $\text{cell\_refs}[c] \gets \text{find\_ref\_cells}(\text{external\_v}, \text{v\_hash})$
\ENDFOR
\RETURN $\mathcal{G} \gets \text{construct\_dag}(\text{cell\_refs})$
\end{algorithmic}
\end{algorithm}
\vspace{-3mm}

\textbf{DAG Construction.}
As shown in Algorithm~\ref{alg:dag}, given notebook cells $\mathcal{C}$, the DAG construction process includes two steps:
\begin{itemize}[leftmargin=10pt]
    \item \textbf{Identify new variables.} For Python cells, we construct an abstract syntax tree (AST) to find \textit{global} variables accessible throughout the notebook (\eg function/class definitions, package imports). We exclude local variables as they are only visible within their scope. For SQL cells, any \texttt{SELECT}'s output is stored in a data variable (\eg a \texttt{DataFrame}) for future use, and thus represents a new variable. Markdown and Chart cells do not produce variables that can be referenced elsewhere, and are thus omitted. We store the variable-cell associations using a hash table.
    \item \textbf{Find referenced cells.} Based on the hash table, we locate each cell's referenced cells by identifying its \textit{external} variables defined in other cells. For Python and SQL cells, this can be easily achieved with ASTs. For Chart cells, the underlying data variable serves as the reference point. As Markdown cells do not associate with any variables, they are excluded from this step. Using the extracted cell references, a DAG of the notebook can be constructed.
\end{itemize}
The DAG keeps updating whenever a cell is created, modified, or deleted, provided that the changes pass the syntax check. This ensures real-time maintenance of cell dependencies.

\textbf{Context Retrieval.}
Based on the cell dependency DAG and an input query, relevant cells are identified through graph traversal. This process supports queries at both \textit{cell-level} and \textit{notebook-level}.
For cell-level queries, the search is initiated within an existing cell, allowing for the straightforward identification of all ancestral cells via the DAG.
Notebook-level queries, conversely, are formulated without specifying an existing cell, which typically rely on LLMs to automatically create new cells. In such cases, we first determine the related data variable either from explicit user input or through LLM prediction. Then, we locate the initial cell $c_s$ where the data variable is defined. To ensure through coverage, all descendant cells of $c_s$ are considered. Additionally, since Markdown cells lack references, our selection is guided by the textual similarity between cell content and the query. This process yields a comprehensive set of relevant cells $\mathcal{C}_r$ for each query.

Subsequently, $\mathcal{C}_r$ is pruned based on task types. Specifically, we employ LLMs to determine the task type contained in the query and identify the involved cell types. For example, in NL2DSCode tasks, only Python cells are considered. Accordingly, we filter out irrelevant cell types, resulting in a pruned set that constitutes the \textit{minimum set} of relevant cells.

We then retrieve the associated information from the shared buffer for each relevant cell generated or altered by agents. The final necessary contexts for the query are determined by combining the retrieved information units with the original relevant cells, thereby providing a concise yet sufficient background for the proxy agent to understand and address the query.
\section{Experiment}

\edit{We evaluate \name \space on both public research benchmarks (§A, §B) and proprietary datasets from Tencent (§C, §D, §E).}

\renewcommand{\arraystretch}{0.88}
\begin{table*}[t]
\caption{Performance of \name \space on research benchmarks
\vspace{-2mm}
}
\label{tab:e2e}
\definecolor{ipv}{RGB}{0, 128, 0}
\definecolor{dcs}{RGB}{128, 0, 0}
\centering
\begin{threeparttable}
\setlength\tabcolsep{2.6pt}{
\begin{tabular}{c|c|c|c|cccc}
\toprule
\textbf{BI Stage} & \textbf{Task} & \textbf{Benchmark}  & \textbf{Metric}  & \multicolumn{4}{c}{\textbf{Method \& Performance}} \\ 
\midrule
\multirow{8}{*}{Data Preparation} &
  \multirow{4}{*}{NL2SQL} & 
  \multirow{2}{*}{Spider~\cite{yu2018spider}} &
  \multirow{2}{*}{Execution Accuracy} &
DataLab (Ours) & DAIL-SQL~\cite{10.14778/3641204.3641221} & \edit{PURPLE~\cite{DBLP:conf/icde/RenFHHDHJZYW24}} & \edit{CHESS~\cite{DBLP:journals/corr/abs-2405-16755}} \\
  & & & & 80.70 & 83.60 & \edit{\textbf{87.80}} & \edit{\underline{87.20}} \\ \cline{3-8} 
  & & \multirow{2}{*}{BIRD~\cite{li2024can}} &
  \multirow{2}{*}{Execution Accuracy} &
  DataLab (Ours) & DAIL-SQL~\cite{10.14778/3641204.3641221} & \edit{PURPLE} & \edit{CHESS} \\
  & & & & 61.33 & 57.41 & \edit{\underline{68.12}} & \edit{\textbf{68.31}}   \\ \cline{2-8} 
  &
  \multirow{4}{*}{NL2DSCode} &
  \multirow{2}{*}{DS-1000~\cite{lai2023ds}} &
  \multirow{2}{*}{Pass Rate} &
  DataLab (Ours) & CoML~\cite{zhang2023mlcopilot} & Code Interpreter~\cite{cia} & \edit{Open Interpreter~\cite{oia}} \\
  & & & & \textbf{53.80} & {44.20} & \underline{51.60} & \edit{50.50} \\ \cline{3-8} 
  & & \multirow{2}{*}{DSEval~\cite{zhang2024benchmarking}} &
  \multirow{2}{*}{Pass Rate} &
  DataLab (Ours) & CoML & Code Interpreter & \edit{Open Interpreter} \\
  & & & & \textbf{80.99} & 71.90 & \underline{80.58} & \edit{78.10} \\
\midrule
\multirow{6}{*}{Data Analysis} &
  \multirow{6}{*}{NL2Insight} & 
      \multirow{2}{*}{DABench~\cite{DBLP:conf/icml/HuZWCM0WSXZCY0K24}} &
      \multirow{2}{*}{Accuracy} &
      DataLab (Ours) & AutoGen~\cite{DBLP:journals/corr/abs-2308-08155} & AgentPoirot~\cite{sahu2024insightbench} & - \\
      & & & & \underline{75.10} & {71.48} & \textbf{75.88} & - \\ \cline{3-8} 
      & &
      \multirow{4}{*}{InsightBench~\cite{sahu2024insightbench}} &
      \multirow{2}{*}{LLaMA-3-Eval} &
      DataLab (Ours) & AutoGen & AgentPoirot & - \\
      & & & & \textbf{0.37} & {0.31} & \underline{0.35} & - \\ \cline{4-8} 
      & & &
      \multirow{2}{*}{ROUGE-1} &
      DataLab (Ours) & AutoGen & AgentPoirot & - \\
      & & & & \underline{0.33} & {0.28} & \textbf{0.35} & - \\
\midrule
\multirow{6}{*}{Data Visualization} &
  \multirow{6}{*}{NL2VIS} & 
      \multirow{2}{*}{nvBench~\cite{luo2021synthesizing}} &
      \multirow{2}{*}{Execution Accuracy} &
      DataLab (Ours) & LIDA~\cite{dibia-2023-lida} & Chat2Vis~\cite{10121440} & \edit{CoML4VIS~\cite{chen2024viseval}} \\
      & & & & \underline{53.90} & \textbf{54.71} & {53.83} & \edit{51.12} \\ \cline{3-8} 
      & &
      \multirow{4}{*}{VisEval~\cite{chen2024viseval}} &
      \multirow{2}{*}{Pass Rate} &
      DataLab (Ours) & LIDA & Chat2Vis & \edit{CoML4VIS} \\
      & & & & \textbf{75.99} & 74.66 & {71.91} & \edit{\underline{75.27}} \\ \cline{4-8} 
      & & &
      \multirow{2}{*}{Readability Score} &
      DataLab (Ours) & LIDA & Chat2Vis & \edit{CoML4VIS} \\
      & & & & 3.73 & \underline{3.77} & {3.70} & \edit{\textbf{3.80}} \\
\bottomrule
\end{tabular}}
\begin{tablenotes}[para,flushleft]
\fontsize{6pt}{6pt}\selectfont
\item \edit{NOTE: The best method is marked in \textbf{bold}, while the second-best method is marked with \underline{underlines}.}
\end{tablenotes}
\end{threeparttable}
\vspace{-7mm}
\end{table*}

\subsection{End-to-End Performance}
\label{e2e}

To demonstrate the capabilities of \name \space as a unified BI platform, we first compare its end-to-end performance with SOTA LLM-based baselines on four typical BI tasks.
\edit{For fair comparison, the baselines employed are all prompt- or agent-based methods without any pre-training or supervised fine-tuning, similar to \name's agent framework.}
We utilize \textbf{GPT-4}~\cite{DBLP:journals/corr/abs-2303-08774} as the foundation model for all methods.

\subsubsection{Settings}
The \textbf{NL2SQL} task converts natural language to SQL queries, typically marking the start of a BI workflow. We use two benchmarks (\ie Spider~\cite{yu2018spider} and BIRD~\cite{li2024can}) and \edit{compare with three baselines (\ie DAIL-SQL~\cite{10.14778/3641204.3641221}, PURPLE~\cite{DBLP:conf/icde/RenFHHDHJZYW24}, and CHESS~\cite{DBLP:journals/corr/abs-2405-16755}}). We use \textit{Execution Accuracy (EX)} as the evaluation metric, which measures the execution equivalence of the generated SQL queries with ground truth.

The \textbf{NL2DSCode} task converts natural language to data science code using Python libraries like \texttt{NumPy} and \texttt{Pandas}, which happens frequently throughout the BI workflow. We use two benchmarks (\ie DS-1000~\cite{lai2023ds} and DSEval~\footnote{We only evaluate DSEval-LeetCode and -SO due to implementation issues.}~\cite{zhang2024benchmarking}) and \edit{compare with three baselines (\ie CoML~\cite{zhang2023mlcopilot}, Code Interpreter~\cite{cia}, and Open Interpreter~\cite{oia}}). \textit{Pass Rate} is used as the evaluation metric, which divides the number of passed problems by all problems.

The \textbf{NL2VIS} task converts natural language to data visualizations based on either Python libraries like \texttt{Matplotlib} or visualization grammars like \texttt{Vega-Lite}. We use two benchmarks (\ie nvBench~\cite{luo2021synthesizing} and VisEval~\cite{chen2024viseval}) and \edit{compare with three baselines (\ie LIDA~\cite{dibia-2023-lida}, Chat2Vis~\cite{10121440}, and CoML4VIS~\cite{chen2024viseval}})~\footnote{As some baselines lack support for NL queries related to multiple data tables, we only evaluate on single-table queries for fair comparison.}. For nvBench, we use the \textit{EX} metric for evaluation, which measures the equivalence of the generated visualizations with the ground truth based on the presented data values and chart types~\cite{DBLP:journals/corr/abs-2401-11255}. For VisEval, we use the \textit{Pass Rate} metric to measure the ratio of valid or legal results divided by all queries, and the \textit{Readability Score} judged by GPT-4V(ision)~\cite{DBLP:journals/corr/abs-2309-17421} to measure the overall quality of the generated visualizations~\cite{chen2024viseval}.

The \textbf{NL2Insight} task converts analysis goals to data insights in an end-to-end manner, which requires LLMs' comprehensive problem-solving abilities. We use two benchmarks (\ie InfiAgent-DABench~\cite{DBLP:conf/icml/HuZWCM0WSXZCY0K24} and InsightBench~\cite{sahu2024insightbench}) and compare with two baselines (\ie AutoGen~\cite{DBLP:journals/corr/abs-2308-08155} and AgentPoirot~\cite{sahu2024insightbench}). For InfiAgent-DABench, we calculate the \textit{Accuracy} of problems with correct answers to all problems. For InsightBench, we use the summary-level \textit{LLaMA-3-Eval} and \textit{ROUGE-1} scores as the evaluation metrics, which measure the alignment of the generated insights against the ground truth based on LLaMA-3 judgment and unigram overlap, respectively~\cite{sahu2024insightbench}.

\subsubsection{Results}
\edit{
As shown in Table~\ref{tab:e2e}, \name \space achieves comparable performance with the SOTA LLM-based baselines on all four BI tasks. These baselines primarily focus on \textit{single} tasks and exhibit the issues discussed in Section~\ref{sec:intro}, such as lacking domain knowledge incorporation or unstructured communication.
Specifically, \name \space outperforms all baselines on benchmarks including DS-1000, DSEval, InsightBench, and VisEval, spreading over each critical BI stage.
While certain baselines excel in \textit{individual} tasks (\eg NL2SQL), the primary focus of \name \space is to unify the BI workflow with a \textit{single} LLM-based framework, maintaining satisfactory performance across \textit{various} tasks. This unification is particularly beneficial for real-world scenarios requiring multi-task coordination.
Moreover, many agent-based SOTA approaches (\eg Data Interpreter~\cite{DBLP:journals/corr/abs-2402-18679}) that follow the reasoning and acting (ReAct) paradigm~\cite{yao2023react} can be integrated into our framework through DAG-based agent workflows (Section~\ref{sec:overview}). This extensibility further enhances \name's practical applicability.
}

For tasks that require the generation of symbolic languages (\eg NL2DSCode, NL2VIS), \name \space consistently performs well primarily due to the intermediate DSL specifications generated by our \textit{Domain Knowledge Incorporation} module. Although most research benchmarks lack the necessary information for extracting table/column knowledge, \name \space adopts a meticulously-designed data profiling strategy as an alternative (Section~\ref{sec:know_utilize}) to fully utilize the provided data schema, enabling LLMs to correctly identify and associate the semantic relationships between data columns and NL queries, which are crucial to generate high-quality DSLs. Consequently, compared to merely feeding the original pure NL queries, these intermediate DSLs can significantly improve LLM-based agents' performance on generating higher-level languages like SQL queries, Python code, or Vega-Lite specifications, as also shown in previous works~\cite{DBLP:conf/nips/WangW0CSK23}.

For more complex tasks (\eg NL2Insight) that typically require multi-step reasoning and/or the collaboration of multiple agents, \name \space also achieves a satisfactory performance. Notably, it outperforms AutoGen, a popular multi-agent framework, by up to 5.06\% on DABench and 19.35\% on InsightBench. This performance gain can be attributed to two key factors: the agents' improved understanding of the involved datasets due to data profiling and the incorporation of our structured communication mechanism. This mechanism standardizes inter-agent information sharing, enabling a more comprehensive and thorough insight discovery process, especially when provided with high-level analytical objectives.

\vspace{-1mm}
\subsection{Sensitivity Analysis}
\label{sens}


\begin{figure}[t]
    \centering
    \includegraphics[width=\linewidth]{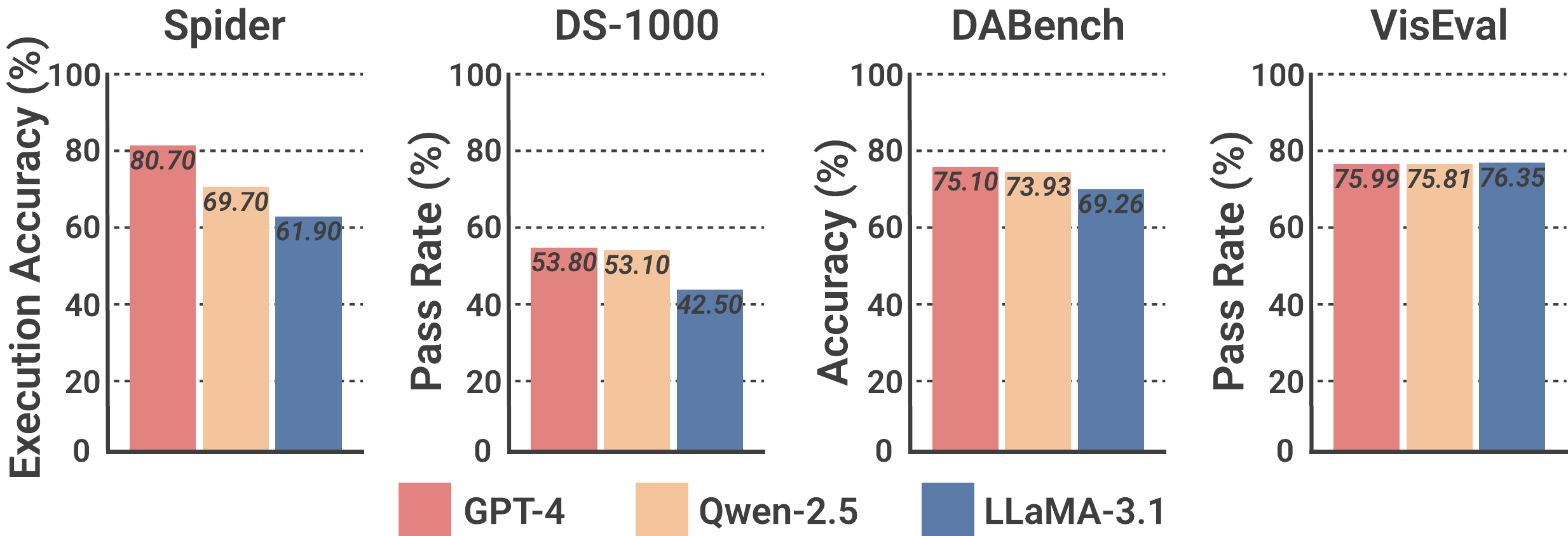}
    \vspace{-7.5mm}
    \caption{Performance of \name \space using various underlying LLMs.}
    \label{fig:sens}
    \vspace{-7mm}
\end{figure}

To evaluate \name's robustness, we experiment with both closed- and open-source LLMs (\ie \textbf{GPT-4}, \textbf{Qwen-2.5}~\cite{qwen2.5}, and \textbf{LLaMA-3.1}~\cite{DBLP:journals/corr/abs-2407-21783}) on the above tasks using benchmarks including Spider, DS-1000, DABench, and VisEval.

As shown in Figure~\ref{fig:sens}, \name \space consistently achieves satisfactory performance on all tasks, albeit with some sensitivity to the underlying LLMs. Proprietary models like GPT-4 typically exhibit superior instruction following and code generation abilities, surpassing open-source models like Qwen-2.5 and LLaMA-3.1. For code-intensive tasks like NL2DSCode and NL2Insight, LLaMA-3.1 experiences notable performance drops, especially on DS-1000, due to its relatively weaker code generation capabilities. We further evaluate DS-1000 using vanilla LLaMA-3.1 and achieve a pass rate of 36.90\% (lower than 42.50\% when integrated with \name). Another interesting fact is that, all three LLMs perform similarly on VisEval, with LLaMA-3.1 surprisingly being the best. These findings indicate that \name \space maintains a consistent performance across tasks, despite variations in LLMs, attributed to our data profiling and communication mechanisms. The data profiling mechanism enhances agents' understanding of input data, while the inter-agent communication module enables efficient error handling and iterative refinement, leading to overall performance improvements.

\subsection{Effect of Domain Knowledge Incorporation}
\label{dk}

\subsubsection{Knowledge Generation}
As described in Section~\ref{sec:know_gen}, this module aims to automatically generate knowledge components of data tables and columns. Deployed at Tencent for one month, \textbf{2,426} databases, \textbf{262,041} tables, and \textbf{2,708,884} columns (averaging 10.3 columns per table) have been successfully processed for knowledge generation, with an average time cost of 45.2 seconds per table. These statistics exhibit the practical application of our approach at a large enterprise.

To evaluate the quality of the generated knowledge, we collect a real-world dataset comprising 50 tables and 629 columns. Each table and column is annotated by domain experts for a ground truth of their semantic meanings. We then compare the \textit{Sentence Embedding Similarity (SES)} between the generated descriptions and the ground truth using M3-Embedding~\cite{DBLP:conf/acl/ChenXZLLL24}, with 1 being identical and 0 being irrelevant. Results show that, the average \textit{SES} scores are \textbf{0.712} (60\% above 0.7) for tables and \textbf{0.677} (53\% above 0.7) for columns, indicating the practical utility of the generated knowledge.

Overall, the real-world deployment and quality evaluation demonstrate the efficiency and effectiveness of the knowledge generation process of this module in practical settings.

\subsubsection{Downstream Tasks}
To assess the real-world impact of this module, we evaluate the following downstream tasks:
\begin{itemize}[leftmargin=10pt]
    \item The \textbf{Schema Linking} task seeks to select relevant tables and columns from the database schema based on NL queries, providing a basis for further analysis~\cite{DBLP:conf/emnlp/LeiWMGLKC20}. It requires LLMs to precisely capture the semantic relationships between user input and elements of the schema.
    \item The \textbf{NL2DSL} task converts NL queries to DSLs, which have been commonly adopted in commercial BI platforms and are crucial for many downstream tasks~\cite{DBLP:journals/cl/PopovicLDD15, su2024tablegpt2largemultimodalmodel}. In \name, DSLs are used as intermediates for generating SQL queries, Python code, and visualizations.
\end{itemize}

Due to common issues like ambiguities and jargon in real-world BI scenarios, both tasks require LLMs' deep understanding of domain knowledge. For Schema Linking, we collect a real-world dataset comprising 439 query-table-column pairs, and use \textit{Recall @5} for evaluation. For NL2DSL, we compile another dataset comprising 326 query-DSL pairs, and measure the overall \textit{Accuracy}. We then employ this module to generate knowledge for each involved table and column. For comparison, we design the following three experiment settings:
\begin{itemize}[leftmargin=10pt]
    \item \textbf{S1 (w/o knowledge)}: This setting provides NL queries along with a brief data schema generated by \texttt{Pandas}, but no additional knowledge, serving as a baseline. It is commonly adopted by most existing LLM-based agents.
    \item \textbf{S2 (w/ partial knowledge)}: Compared to S1, this setting additionally provides the generated \textit{description}, \textit{usage}, and \textit{tags} of data tables and columns. It accounts for almost all successful cases in our practical deployment.
    \item \textbf{S3 (w/ all knowledge)}: Compared to S2, this setting further provides all generated knowledge of data tables and columns (see Section~\ref{sec:know_gen}). It accounts for approximately 40\% successful cases in our practical deployment.
\end{itemize}

As shown in Table~\ref{tab:dk}, \name's performance on both tasks improves significantly when provided with enterprise-specific knowledge. Specifically, the \textit{Recall @5} of Schema Linking increases by \textbf{38.47\%}, and the \textit{Accuracy} of NL2DSL improves by up to \textbf{58.58\%}. Even with only partial knowledge (S2), the performance still exhibits a significant increase of 30.77\% for Schema Linking and 29.14\% for NL2DSL compared to the baseline (S1). During our deployment at Tencent, we observe that many real-world business tables lack sufficient information required for generating comprehensive knowledge, often limited to table and column \textit{descriptions}, \textit{usage}, and \textit{tags}. While understanding the semantic meanings of ambiguous table/column names can largely enhance LLMs' performance, the absence of other knowledge - especially \textit{calculation logic} of derived columns for NL2DSL - can impede their capabilities in certain scenarios. This explains the performance difference between S2 and S3. Despite this, the promising results of S2 guarantee a minimum acceptable level of performance, demonstrating the module's effectiveness and robustness for downstream tasks in real-world BI scenarios.

\subsection{Effect of Inter-Agent Communication}
\label{exp:ablation_inter_agent_communication}
We experiment with a complex BI scenario that involves multiple tasks performed by distinct agents: NL2SQL, NL2DSCode, NL2VIS, Anomaly Detection, Causal Analysis, and Time Series Forecasting. We compile a dataset from practical settings at Tencent, consisting of 2 databases, 10 tables, and 111 columns. For each table, we meticulously design 10 complex questions derived from real-world business queries, totaling to 100 samples. Each question requires multi-step reasoning and multi-agent collaboration, ensuring a rigorous evaluation of our inter-agent communication mechanism.

We evaluate this module's efficiency and effectiveness by respectively calculating the \textit{Success Rate} and \textit{Accuracy} of the agents' responses across all questions. The \textit{Success Rate} measures the ratio of questions that can be successfully solved within a maximum of 5 calls per agent, while the \textit{Accuracy} measures the ratio of correct answers among all questions.
For comparison, we employ three experiment settings:
\begin{itemize}[leftmargin=10pt]
    \item \textbf{S1 (w/o FSM)}~\cite{hong2023metagpt4}: This setting removes the FSM-based information sharing protocol. Therefore, each agent receives \emph{all} information from the shared buffer.
    \item \textbf{S2 (w/o information formatting)}~\cite{DBLP:journals/corr/abs-2308-08155}: This setting removes the information format structure and adopts \emph{pure natural language} for inter-agent communication.
    \item \textbf{S3 (w/ both)}: This setting keeps both techniques.
\end{itemize}

\begin{table}[t]
\caption{Ablation study on Domain Knowledge Incorporation}
\vspace{-2mm}
\label{tab:dk}
\begin{tabularx}{\linewidth}{
    >{\centering\arraybackslash\hsize=2.2\hsize}X|
    >{\centering\arraybackslash\hsize=0.6\hsize}X|
    >{\centering\arraybackslash\hsize=0.6\hsize}X|
    >{\centering\arraybackslash\hsize=0.6\hsize}X
}
\toprule
\textbf{Task / Metric}        & \textbf{S1} & \textbf{S2} & \textbf{S3}    \\ \midrule
Schema Linking / Recall @5 (\%) & 41.02 & 71.79 & \textbf{79.49} \\ 
NL2DSL / Accuracy (\%)           & 32.52       & 61.66       & \textbf{91.10} \\ \bottomrule
\end{tabularx}
\vspace{-4mm}
\end{table}

\begin{table}[t]
\caption{Ablation study on Inter-Agent Communication}
\vspace{-2mm}
\label{tab:com}
\begin{tabularx}{\linewidth}{
    >{\centering\arraybackslash\hsize=1.5\hsize}X|
    >{\centering\arraybackslash\hsize=0.8333\hsize}X|
    >{\centering\arraybackslash\hsize=0.8333\hsize}X|
    >{\centering\arraybackslash\hsize=0.8333\hsize}X
}
\toprule
\textbf{Metric}     & \textbf{S1} & \textbf{S2} & \textbf{S3}    \\ \midrule
Success Rate (\%) & 73.00  & 85.00 & \textbf{92.00} \\ 
Accuracy (\%) &  56.00  &   79.00    & \textbf{84.00} \\ \bottomrule
\end{tabularx}
\vspace{-7mm}
\end{table}

As illustrated in Table~\ref{tab:com}, \name's performance on complex BI tasks improves by \textbf{19.00\%} in \textit{Success Rate} and \textbf{28.00\%} in \textit{Accuracy} with our inter-agent communication mechanism. Without the FSM-based information sharing protocol (S1), performance significantly degrades. Error analysis reveals that most failures involve more than 3 agents, resulting in overwhelming and irrelevant information that hinders LLMs' reasoning, thereby leading to incorrect outputs~\cite{DBLP:conf/icml/ShiCMSDCSZ23}. Additionally, the absence of the information format structure (S2) leads to a 7\% decrease in \textit{Success Rate} and a 5\% drop in \textit{Accuracy}, highlighting the importance of structured prompts in enhancing LLM comprehension and reducing information sharing ambiguities. This is critical in BI scenarios where complex tasks often require iterative error handling for data processing and structured summaries for lengthy outputs.

\subsection{Effect of Cell-based Context Management}

\subsubsection{DAG Construction}
To evaluate the efficiency of the DAG construction process, we collect 50 \name \space notebooks containing multi-language cells from practical settings, with cell counts ranging from 2 to 49. We measure the \textit{Time Cost} of DAG construction both at notebook-level and cell-level. The initial construction encompasses all cells upon notebook opening, whereas subsequent updates generally involve a single cell. Our goal is to ensure a reasonable cold-start time while maintaining real-time responsiveness for subsequent updates.

\begin{figure}[t]
    \centering
    \includegraphics[width=\linewidth]{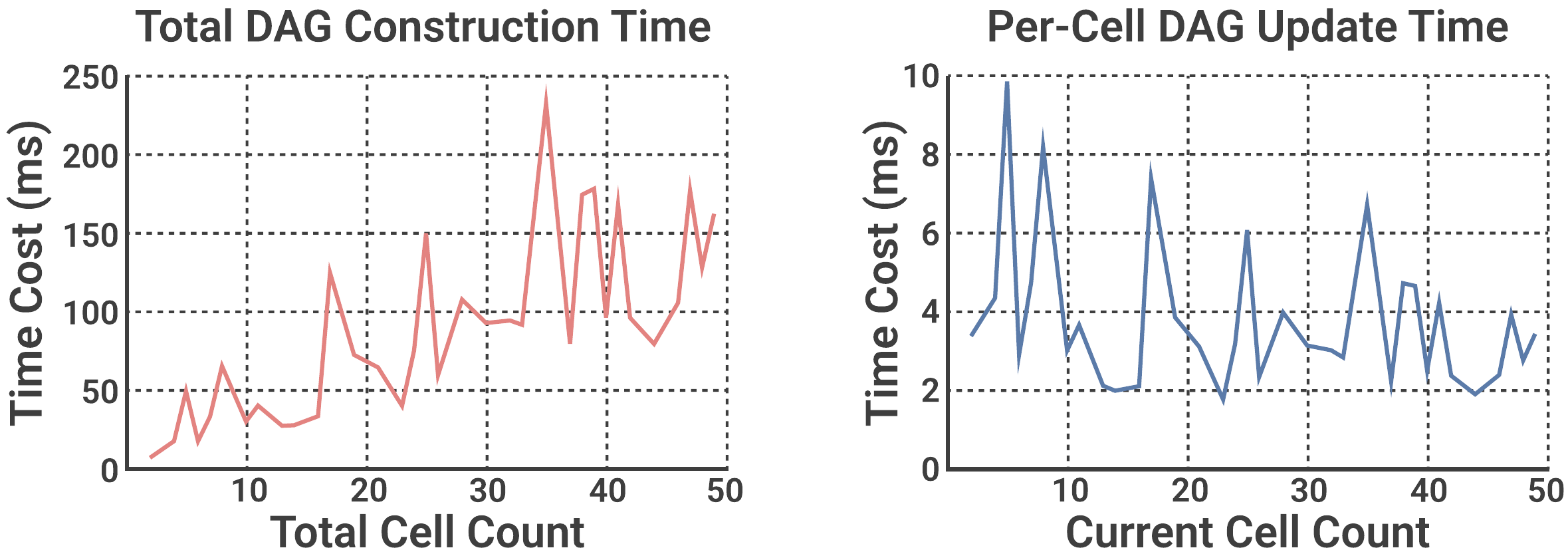}
    \vspace{-6mm}
    \caption{Time cost of DAG construction.}
    \label{fig:dag}
    \vspace{-4mm}
\end{figure}

As shown in Figure~\ref{fig:dag}, DAG construction and updating maintain low time costs, at less than \textbf{250} and \textbf{10} \textbf{milliseconds}, respectively. The total time for DAG construction increases with cell count, reaching a maximum of 232.22 milliseconds for 35 cells. In contrast, the per-cell time for DAG updating averages to 3.78 milliseconds, peaking at 9.84 milliseconds for 5 cells. Time costs are affected not only by cell count but also by lines of code, accounting for observed fluctuations. Given that a typical \name \space notebook contains fewer than 50 cells, these results demonstrate the efficiency of DAG construction.

\begin{table}[t]
\caption{Ablation study on Cell-based Context Management}
\vspace{-2mm}
\label{tab:context}
\begin{tabularx}{\linewidth}{
>{\centering\arraybackslash\hsize=1.4\hsize}X|
>{\centering\arraybackslash\hsize=0.8\hsize}X|
>{\centering\arraybackslash\hsize=0.8\hsize}X
}
\toprule
\textbf{Metric}        & \textbf{S1} & \textbf{S2}    \\ \midrule
Accuracy (\%) & \textbf{86.67} & 82.00 \\
Token Cost per Query (K) & 10.69 & \textbf{4.10}       \\ \bottomrule
\end{tabularx}
\vspace{-7mm}
\end{table}

\subsubsection{Task Completion}
For each notebook in our collected dataset, we derive 3 real-world user queries, which involve NL2SQL, NL2DSCode, and NL2VIS tasks, totaling to 150 samples. We evaluate this module's performance and cost-effectiveness using two metrics: \textit{Accuracy} and \textit{Token Cost per Query}. For comparison, we conduct an ablation study with two experiment settings: \textbf{S1 (w/o DAG)} and \textbf{S2 (w/ DAG)}.

As illustrated in Table~\ref{tab:context}, \name \space achieves a satisfactory \textit{Accuracy} under both settings. Further analysis reveals that certain Markdown cells may contain critical information for task completion, which are occasionally failed to retrieve by our context retrieval mechanism due to limitations of embedding similarity~\cite{DBLP:conf/www/SteckEK24}. This accounts for the slight 4.67\% drop in \textit{Accuracy} under S2. However, S2 significantly reduces the \textit{Token Cost per Query} by \textbf{61.65\%} compared to S1, \edit{saving approximately \$65.90 for every 1,000 queries using GPT-4.}
This is achieved by identifying the \textit{minimum set} of relevant cells based on DAGs. These results demonstrate this module's cost-effectiveness while maintaining acceptable performance.
\vspace{-5mm}
\section{\edit{Real-World Applications}}
\edit{
\name \space has been effectively deployed on Tencent TEG's Big Data Platform, achieving an average of 2,093 monthly registered users, 10,900 monthly API calls, and 319 weekly active users over a three-month period. It significantly improves the efficiency of data professionals by integrating LLM-powered BI tasks into a unified notebook interface. Below, we highlight three practical use cases to illustrate its benefits.
}

\edit{
\vspace{-5mm}
\begin{itemize}[leftmargin=10pt]
    \item Fiona is a data engineer who regularly prepares data for product managers. Previously, she typically adds domain knowledge to prompts manually to enhance performance. With \name's \textit{Domain Knowledge Incorporation} module, she is surprised to discover that it can inherently understand her requirements even with ambiguous column names.
    \item Henry, a data scientist specializing in recommendation algorithms, values \name \space for generating and displaying Python code directly in notebook cells. It makes editing easier and eliminates the need to repetitively copy-paste between chat-based LLM interfaces and traditional notebooks.
    \item Jerry is a data analyst who provides data reports for stakeholders. Using \name, he cuts the time to create data visualizations by 87\% (from 15 minutes to 2 minutes). He highly appreciates \name's Chart cells, which enable GUI-based customization of automatically generated visualizations.
\end{itemize}
}

\vspace{-0.5mm}
\edit{
Additionally, users agree that working on \name \space facilitates communication efficiency with colleagues. For example, they can leave comments next to cells to quickly notify collaborators about their progress, avoiding the need to tediously share screenshots or code snippets as they did previously.
}
\vspace{-0.1mm}
\section{Related Work}

\textbf{Business Intelligence Platforms.}
BI platforms support users in analyzing business data for decision-making~\cite{meduri2021birecguideddataanalysis, DBLP:journals/corr/abs-2405-00527}.
Representatives like Tableau~\cite{tableau}, Power BI~\cite{powerbi}, and Databricks~\cite{databricks} provide GUIs to support user interactions for data transformation and dashboard generation.
These platforms also integrate natural language interfaces~\cite{DBLP:conf/ieeevast/ToryS19, DBLP:journals/tvcg/FengWPWRLYMQC24} to lower the burden of manual operation.
Quamar \textit{et al.}~\cite{10.14778/3415478.3415557} proposed an ontology-based method based on business models to provide semantic information and reasoning capability for query interpretation.
The emergence of LLMs has further enhanced the domain knowledge integration and visualization generation abilities of BI platforms \cite{DBLP:conf/sigmod/MiaoJ024, DBLP:journals/corr/abs-2402-02643}.
\edit{
Unlike existing tools that focus on visualization (\eg Tableau, Power BI) or are optimized for engineering workflows (\eg Databricks), \name \space provides a unified platform to satisfy various BI stages (\ie data preparation, analysis, and visualization) and data roles (\ie data engineers, scientists, and analysts) through a one-stop LLM-based agent framework.
}

\textbf{LLM-based Data Analysis.}
LLMs have shown remarkable abilities in semantic understanding and logical reasoning, enabling complex data analysis through conversational interfaces~\cite{DBLP:journals/corr/abs-2404-01644, DBLP:conf/sigmod/ChenLJSFF0024, DBLP:journals/corr/abs-2310-16164}.
For example, Table-GPT~\cite{DBLP:journals/pacmmod/LiHYCGZF0C24} fine-tunes LLMs on synthesized table-task data to enhance their table-understanding abilities. Chat2Query~\cite{DBLP:conf/icde/ZhuCNNXHWMWZTL24} decomposes NL2SQL tasks into multiple steps to improve generation quality. InsightPilot~\cite{DBLP:conf/emnlp/00040WH023} automates the discovery of data insights and synthesizes them into high-level overviews.
\edit{
Moreover, Chat2Data~\cite{DBLP:journals/pvldb/ZhaoZL24} leverages domain knowledge based on vector databases to mitigate LLMs' hallucination issues, while our work uniquely facilitates domain-specific data analysis by automatically extracting knowledge from enterprise scripts and data lineage information, avoiding manual curation. Unlike existing works that focus on end-to-end results, we introduce a novel notebook interface that supports flexible human intervention - allowing refinement of intermediate SQL, Python, or charts - while unifying tasks fragmented across prior task-specific approaches like LIDA~\cite{dibia-2023-lida} and PURPLE~\cite{DBLP:conf/icde/RenFHHDHJZYW24}.
}

\section{Conclusion}

This paper introduces \name, a unified BI platform that combines an LLM-based agent framework with a notebook interface. \name \space features a domain knowledge incorporation module, an inter-agent communication mechanism, and a cell-based context management strategy. These components enable seamless integration of LLM assistance with user customization, making \name \space well-suited for practical BI scenarios.
\name \space has proven effective on both research benchmarks and real-world business datasets from Tencent.


\bibliographystyle{IEEEtran}
\bibliography{main}

\end{document}